\documentclass[amsmath, amsfonts, showpacs, superscriptaddress, twocolumn, prb]{revtex4}
\usepackage{graphicx}
\usepackage{epsfig}
\usepackage{bm}
\usepackage{dcolumn}
\usepackage{amsmath}
\usepackage{amssymb}
\usepackage{color}
\usepackage{stackrel}
\usepackage{accents}
\usepackage{latexsym}
\usepackage{verbatim}
\usepackage{hyperref}

\newcommand{\bp}{\mathbf p}

\newcommand{\br}{\mathbf r}
\newcommand{\bn}{\mathbf n}
\newcommand{\bR}{\mathbf R}
\newcommand{\dg}{^\dagger}
\newcommand{\up}{\uparrow}
\newcommand{\dn}{\downarrow}

\begin{document}

\title{Magnetic penetration depth in disordered iron-based superconductors}

\author{M.~Dzero}
\affiliation{Department of Physics, Kent State University, Kent, Ohio 44242, USA}
\affiliation{Max Planck Institute for the Physics of Complex Systems, N\"{o}thnitzer str. 38, 01187 Dresden, Germany}

\author{M.~Khodas}
\affiliation{Department of Physics and Astronomy, University of Iowa, Iowa City, Iowa 52242, USA}
\affiliation{Racah Institute of Physics, Hebrew University of Jerusalem, Jerusalem 91904, Israel}

\author{A.~D.~Klironomos}
\affiliation{American Physical Society, 1 Research Road, Ridge, New York 11961-9000}

\author{M.~G.~Vavilov}
\affiliation{Department of Physics, University of Wisconsin-Madison, Madison, Wisconsin 53706, USA}

\author{A.~Levchenko}
\affiliation{Department of Physics, University of Wisconsin-Madison, Madison, Wisconsin 53706, USA}
\affiliation{Institut fur Nanotechnologie, Karlsruhe Institute of Technology, 76021 Karlsruhe, Germany}

\begin{abstract}
We study the effect of disorder on the London penetration depth in iron-based superconductors.
The theory is based on a two-band model with quasi-two-dimensional Fermi surfaces, which allows for the coexistence region in the phase diagram between magnetic and superconducting states in the presence of intraband and interband  scattering. Within the quasiclassical approximation we derive and solve Eilenberger's equations, which include a weak external magnetic field, and provide analytical expressions for the penetration depth in the various limiting cases. A complete numerical analysis of the doping and temperature dependence of the London penetration depth reveals the crucial effect of disorder scattering, which is especially pronounced in the coexistence phase. The experimental implications of our results are discussed.
\end{abstract}

\date{June 16, 2015}


\maketitle

\section{Introduction}

Measurements of the magnetic penetration depth $\lambda_L$ as a function of temperature, doping, magnetic field, and crystal orientation provide invaluable information about the nature of superconductivity and the symmetry of the underlying order parameter (see e.g. reviews [\onlinecite{Review-1,Review-2}]). In a single-component clean $s$-wave BCS superconductor, with order parameter $\Delta$ and a fully gapped Fermi surface, the low-temperature behavior of the London penetration depth $\delta\lambda_L(T)=\lambda_L(T)-\lambda_L(0)$ shows an exponential decrease with temperature
\begin{equation}\label{LondonSwave}
\frac{\delta\lambda_L(T)}{\lambda_L(0)}\propto\sqrt{\frac{\Delta}{T}} e^{-\Delta/T}.
\end{equation}
Even though nonmagnetic disorder does not directly affect $\Delta$, it does modify the value of $\lambda_L$, which becomes $\lambda_L^{-2}(T)\propto\Delta(T)\sigma\tanh[{\Delta(T)}/{2T}]$, where $\sigma$ is the normal-state conductivity.~\cite{AGD,Kogan}

In contrast to the $s$-wave case, $d$-wave symmetry of the order parameter with nodes on the Fermi surface translates to power-law temperature dependence for the penetration depth \cite{Yip-Sauls}
\begin{equation}\label{LondonDwave}
\frac{\delta\lambda_L(T)}{\lambda_L(0)}\propto\frac{T}{\Delta}.
\end{equation}
The power exponent of the low-temperature behavior is very sensitive to disorder scattering, such that $\delta\lambda_L$ crosses over to quadratic behavior, $\delta\lambda_L(T)/\lambda_L(0)\simeq (T/\Delta)^2$, below a certain temperature scale $T^*$, which is determined by the concentration of strong scatterers.~\cite{Hirschfeld,Kosztin}

The dependence of the penetration depth on various parameters in the case of iron-pnictide superconductors (FeSCs) is of special interest. These materials have multiple Fermi pockets with electron-like and hole-like dispersion of carriers. Because of a delicate interplay between interactions in various pairing channels, superconductivity in FeSCs emerges in close proximity to a spin-density-wave (SDW) order, and the superconducting (SC) critical temperature $T_c$ has a dome-shaped dependence on doping, with the $T_c$ maximum near the onset of SDW order.~\cite{PD-Exp-1,PD-Exp-2,PD-Exp-3,PD-Exp-4}
It has been proposed~\cite{Review-3} that superconductivity in FeSCs is unconventional, with the order parameter having opposite signs on different Fermi sheets, and named $s^{\pm}$ symmetry. The latter emerges because SDW fluctuations increase interpocket interaction, which is attractive for $s^{\pm}$ gap symmetry, to a level where it overcomes intrapocket repulsion. Likewise, SC fluctuations tend to increase the tendency towards SDW.

The emergent complexity of FeSCs with competing superconducting and magnetic instabilities, which may coexist in a certain region of the phase diagram,~\cite{VVC-PRB10,FS-PRB10} leads to peculiar dependencies of the penetration depth. Early experiments in 122-materials, Co- and K-doped BaFe$_2$As$_2$, revealed that down to the lowest temperatures and in a wide range of dopings the $T$-dependence of $\delta\lambda_L$ can be systematically fitted by $\delta\lambda_L\propto T^2$.~\cite{Gordon-1,Gordon-2,Martin-1} In contrast, in 1111-compounds such as SmFeAsO$_{1-x}$F$_x$~\cite{Malone-1} and PrFeAsO~\cite{Hashimoto-1} the penetration depth has an exponential temperature dependence consistent with a gap without nodes; no appreciable effect of scattering was observed. At the same time, data on another 1111-material, LaFePO~\cite{Fletcher-1} pointed out that $\delta\lambda_L(T)$ varies approximately linearly with $T$, strongly suggesting the presence of gap nodes in this compound. Since these initial reports, the London penetration depth has been measured systematically in a variety of families of iron-pnictides and iron-chalcogenides.~\cite{Hashimoto-2,Hashimoto-3,Gordon-3,Kim-1,Gordon-4,Kim-2,Cho-1,Cho-2,Kim-3,Strehlow,Kim-4} Perhaps the most striking recent observation is a disorder-induced topological change of the superconducting gap structure, as revealed from the low-$T$ behavior of $\delta\lambda_L$ in BaFe$_2$(As$_{1-x}$P$_x$)$_2$.~\cite{Mizukami} Nonmagnetic defects were controllably introduced by electron irradiation, and it was found that the nodal state of P-doped BaFe$_2$As$_2$ changes to a nodeless state with increasing disorder. Moreover, under further irradiation, the gapped state evolves into a different gapless state, thus providing evidence of unconventional sign-changing $s$-wave superconductivity. Such unusual sensitivity of the superconducting gap structure to disorder scattering is a unique characteristic feature of FeSCs.

Theoretical studies of the penetration depth in FeSCs were discussed in Refs.~[\onlinecite{Nagai,Parish,Fernandes,Kuzmanovski}] for clean samples based on the band model. The effects of disorder on the phase diagram, including pair-breaking scattering, and on the penetration depth were investigated in Refs.~[\onlinecite{VVC-PRB09,Bang,Mishra-PRB11,VC-PRB11,FVC,Mishra-PRB13,Hoyer}]. We study the effect of disorder on $\lambda_L$ in a systematic way and analyze its behavior in the part of the phase diagram where the SC and SDW phases coexist. On the technical side, we develop a formalism that enables us to study the doping and temperature evolution of the penetration depth in the whole parameter space of the phase diagram. Recently electron irradiation was used to introduce disorder into FeSC systems in a controlled way.~\cite{Strehlow,Mizukami,Prozorov} Thus, our theory is relevant for the interpretation of existing and future experiments along this exciting direction.

This paper is organized as follows: In Sec.~II we present our model, discuss underlying approximations and assumptions, and we analyze the phase diagram of the FeSC compounds. In Sec.~III we derive and solve quasiclassical Eilenberger equations with emphasis on the coexistence of SC and SDW orders. We then apply that formalism to study the London penetration depth across the whole range of the phase diagram, and at different temperatures. In Sec.~IV we summarize our findings and place our work in the context of future developments.

\section{Model and approximations}

In this section we introduce the minimal model for iron-based superconductors in which doping acts as a source of disorder and produces a region of coexistence between superconductivity and magnetism. Furthermore, right from the outset, we consider the case of nonzero external magnetic field that acts on orbital electron motion, but assumed to be weak enough not to affect spin. We discuss the ground state properties of this model in zero field within the quasiclassical approximation, which we use later to compute the penetration depth across a wide doping range from the coexistence region to the purely superconducting state.

\subsection{Model}
Following the discussion in Refs.~[\onlinecite{VC-PRB11,FVC}], we consider a model with two cylindrical Fermi surfaces. One Fermi surface has electron-type and another one has hole-type excitations. We introduce the following eight-component spinor
\begin{equation}\label{PsiSpinor}
\overline{\Psi}(\br)=(\hat{\psi}_{c}\dg(\br),~\hat{\psi}_{c}(\br),~\hat{\psi}_{f}\dg(\br),~\hat{\psi}_{f}(\br)),
\end{equation}
where $\hat{\psi}_{a}(\br)=(\psi_{a\up}\dg(\br), ~\psi_{a\dn}(\br))$ ($a=c,f$) is a Gor'kov-Nambu spinor, and $\psi_{a\sigma}\dg(\br)$ are the creation operators for the electron ($a$=$f$) and hole ($a$=$c$) fermionic excitations at point $\br$ in real space with a spin component $\sigma=\uparrow\downarrow$. 

The full Hamiltonian for the problem at hand 
\begin{equation}
\mathcal{H}=\frac{1}{2}\sum_{\br\alpha\beta}\overline{\Psi}_{\alpha}(\br)[\hat{H}(\br)]_{\alpha\beta}{\Psi}_{\beta}(\br),
\end{equation}
consists of kinetic part and interactions
\begin{equation}\label{H}
\hat{H}(\br)=\hat{H}_0(\br)+\hat{H}_{\textrm{mf}}(\br).
\end{equation}
In the limit of weak magnetic field the noninteracting Hamiltonian matrix $[\hat{H}_0(\br)]_{\alpha\beta}$ can be compactly written as
\begin{equation}\label{H0ab}
\hat{H}_0(\br)=-\hat{\xi}\hat{\tau}_3\hat{\rho}_3\hat{\sigma}_0+\frac{ie}{mc}{\mathbf A}\cdot{\mbox{\boldmath $\nabla$}}
\hat{\tau}_3\hat{\rho}_0\hat{\sigma}_0.
\end{equation}
Here, $\hat{\tau}_i$, $\hat{\rho}_i$, and $\hat{\sigma}_i$ with $i=0,1,2,3$ are sets of Pauli matrices acting correspondingly in the band, Gor'kov-Nambu, and spin spaces;
$\hat{\tau}_0$,$\hat{\rho}_0$,$\hat{\sigma}_0$ are unit matrices, ${\mathbf A}(\br)$ is the vector potential,
$\hat{\xi}=-\bm{\nabla}^2/{2m}-\mu$, and $\mu$ is the chemical potential.

Interactions between the quasiparticles on the electron- and hole-like Fermi surfaces lead to the development of superconducting and spin-density-wave orders. Within the mean-field theory approximation, the corresponding expression for the interaction part of the model Hamiltonian $\hat{H}_{\mathrm{mf}}$ reads
\begin{eqnarray}\label{Hmf}
&&\hat{H}_{\textrm{mf}}=\hat{H}_\Delta+\hat{H}_M, \\
&&\Hat{H}_\Delta=-\Delta\hat{\tau}_3\hat{\rho}_2\hat{\sigma}_2,\quad\hat{H}_M=\hat{\tau}_1\hat{\rho}_3\bm{M}\cdot\hat{\bm{\sigma}}.
\end{eqnarray}
Here, $\Delta$ is the superconducting order parameter, while $\bm{M}$ is the spin-density-wave order parameter. We explicitly assume that the magnetic field is weak enough so that we can ignore the spatial dependence of both SC and SDW fields [see discussion after Eq.~(\ref{Gperturb}) below].
Also note that $\Delta$ and $\bm{M}$ must be computed self-consistently; we will derive the corresponding equations in what follows. We emphasize that within the model under consideration, we study the case of $s^{\pm}$ pairing, i.e.~the superconducting order parameters on the electron-like and hole-like Fermi surfaces have opposite signs, $\Delta^{(c)}=-\Delta^{(f)}=\Delta$. We also ignore the possible mismatch due to differences in the band occupations and effective masses between the two Fermi surfaces. 

Let us now introduce a disorder potential. In what follows we consider two types of disorder scattering: the first type is intraband disorder with potential $U_0$, which scatters quasiparticles within the same band, while the second type with potential $U_\pi$ accounts for interband scattering. Thus, in the basis (\ref{PsiSpinor}) for disorder potential we write
\begin{equation}\label{Ur}
\hat{U}(\br)=\sum\limits_{i}\left[U_0\hat{\tau}_0\hat{\rho}_3\hat{\sigma}_0+U_\pi \hat{\tau}_1\hat{\rho}_3\hat{\sigma}_0\right]\delta(\br-{\mathbf R}_i),
\end{equation}
where the summation goes over impurity sites. We assume that concentration of impurities is $x_{\textrm{imp}}$.

We will treat the effects of disorder within the self-consistent Born approximation. Specifically, we introduce a single-particle Green's function in the Matsubara representation \cite{AGD} as a solution of the following matrix equations
\begin{equation}\label{HatG}
\begin{split}
&\left[i\hat{\omega}_n-\hat{H}(\br_1)-\hat{\Sigma}_\omega({\mathbf R})\right]\hat{G}(i\omega_n,\br_1,\br_2)=\hat{I}, \\
&\left[-i\hat{\omega}_n-\hat{H}(\br_2)-\hat{\Sigma}_\omega({\mathbf R})\right]\hat{G}(i\omega_n,\br_1,\br_2)=\hat{I},
\end{split}
\end{equation}
where $\hat{\omega}_n=\pi T(2n+1)\hat{\tau}_0\hat{\rho}_0\hat{\sigma}_0$, $T$ is a temperature, $\hat{I}=\hat{\tau}_0\hat{\rho}_0\hat{\sigma}_0\delta(\br_1-\br_2)$, and ${\mathbf R}=(\br_1+\br_2)/2$ is the center-of-mass coordinate. In order to write down an explicit expression for the self-energy $\hat{\Sigma}(i\omega_n,{\mathbf R})$, in addition to the center-of-mass coordinate we introduce the relative coordinate $\br=\br_1-\br_2$, and consider the matrix Green's function (\ref{HatG}) as a function of ${\mathbf R}$ and $\br$. Furthermore, we perform the Fourier transformation with respect to the relative coordinate $\br$, and in what follows we consider the function $\hat{G}(i\omega_n,{\mathbf R},\bp)$. Then, assuming that disorder is uncorrelated, upon averaging over various disorder configurations \cite{AGD} we find the following expression for the self-energy (hereafter $\hbar=c=1$):
\begin{eqnarray}\label{Sigma}
&&\hskip-.25cm\hat{\Sigma}_{\omega}({\mathbf R})=\frac{4\Gamma_0}{\pi\nu}\int\frac{d^2\bp}{(2\pi)^2}\hat{\tau}_0\hat{\rho}_3\hat{\sigma}_0
\hat{G}(i\omega_n,{\mathbf R},\bp)\hat{\tau}_0\hat{\rho}_3\hat{\sigma}_0\nonumber\\
&&\hskip-.25cm+\frac{4\Gamma_\pi}{\pi\nu}\int\frac{d^2\bp}{(2\pi)^2}\hat{\tau}_1\hat{\rho}_3\hat{\sigma}_0
\hat{G}(i\omega_n,{\mathbf R},\bp)\hat{\tau}_1\hat{\rho}_3\hat{\sigma}_0,
\end{eqnarray}
where the cross terms $\propto U_0U_\pi$ vanish; $\nu$ is the single particle density of states, $\Gamma_0=\pi\nu x_{\textrm{imp}}|U_0|^2/4$, and
$\Gamma_{\pi}=\pi\nu x_{\textrm{imp}}|U_\pi|^2/4$. Clearly, the fully self-consistent computation of the order parameters $\Delta$ and $\bm{M}$ along with the self-energy $\hat{\Sigma}_\omega(\mathbf{R})$ is a challenging problem. However, this problem can be solved efficiently using the quasiclassical approach.

\subsection{Quasiclassical approximation}

The quasiclassical approximation is justified when the characteristic quantities for the problem at hand vary significantly on length scales that are much longer than the Fermi wavelength $\lambda_F$. In the context of iron-based superconductors, the quasiclassical approximation works well since both superconducting and magnetic correlation lengths greatly exceed $\lambda_F$.~\cite{VC-PRB11,FVC,Hoyer,Moor}

The central object in the quasiclassical approach is the Eilenberger function~\cite{Eilenberger}
\begin{equation}\label{EilenG}
\hat{\cal G}_{\omega}({\mathbf R},{\mathbf n})=\frac{4i}{\pi\nu}\int\frac{pdp}{2\pi}\hat{\tau}_3\hat{\rho}_3\hat{\sigma}_0\cdot\hat{G}(i\omega_n,{\mathbf R},\bp).
\end{equation}
To derive an equation for the Eilenberger function $\hat{\cal G}_{\omega}({\mathbf R},{\mathbf n})$ in a weak external magnetic field, one needs to eliminate the single-particle dispersion via a series of algebraic manipulations (see Appendix A for details). Taking into account that the relevant values of the quasiparticle momentum $\bp$ are close to the Fermi momentum $p_F$, so that
$\bp/m\approx v_F{\mathbf n}$, we  find the following equation for $\hat{\mathcal{G}}_\omega$:
\begin{eqnarray}\label{EilenFinGRn}
&&\left[i\omega_n\hat{\tau}_3\hat{\rho}_3\hat{\sigma}_0,\hat{\cal G}_\omega({\mathbf R},{\mathbf n})\right]
-\left[\hat{H}_{\textrm{mf}}\hat{\tau}_3\hat{\rho}_3\hat{\sigma}_0,\hat{\cal G}_\omega({\mathbf R},{\mathbf n})\right]\nonumber\\
&&-\left[\hat{\Sigma}_{\omega}({\mathbf R})\hat{\tau}_3\hat{\rho}_3\hat{\sigma}_0,\hat{\cal G}_\omega({\mathbf R},{\mathbf n})\right]\nonumber\\
&&+\left[ev_F\bn\cdot{\mathbf A}(\bR)\hat{\tau}_0\hat{\rho}_3\hat{\sigma}_0,\hat{\cal G}_\omega({\mathbf R},{\mathbf n})\right]\nonumber\\
&&+v_F\bn\cdot(-i{\mbox{\boldmath $\nabla$}_\bR})\hat{\cal G}_\omega({\mathbf R},{\mathbf n})=0,
\end{eqnarray}
where the square brackets denote a commutator. Since we consider the limit of a weak magnetic field, we can look for the solution of this equation by perturbation theory, namely
\begin{equation}\label{Gperturb}
\hat{\cal G}_\omega({\mathbf R},{\mathbf n})=\hat{\cal G}^{(0)}_\omega+\hat{\cal G}_\omega^{(1)}({\mathbf R},{\mathbf n}),
\end{equation}
restricting ourselves to corrections linear in powers of the vector potential ${\mathbf A}$. This is why we could neglect the dependence of $\Delta$ and $\bm{M}$ on $\bR$, since the corrections that render both order parameters spatially inhomogeneous are of the order of $O({\mathbf A}^2)$.\cite{AGD} Next, we discuss the solution of the Eilenberger equation in the spatially homogeneous case.

\begin{figure}
\includegraphics[width=1\linewidth]{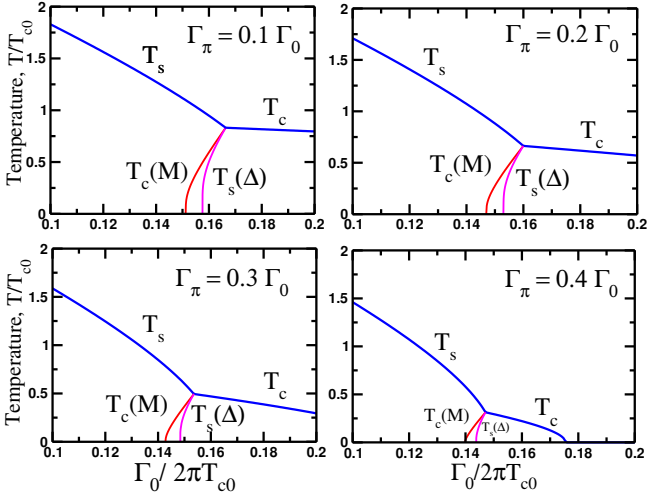}
\caption{\label{Fig1Dia} (Color online) Phase diagrams in the $(\Gamma_0,T)$ plane computed by solving the self-consistency equations (\ref{selfEilen}) and the Eilenberger equations (\ref{FinEilen}) for various ratios between intraband and interband scattering rates. $T_{c0}$ denotes the superconducting critical temperature in a clean system. All plots are obtained assuming $T_{s0}=3T_{c0}$, and $T_{s0}$ is the critical temperature for the SDW state.}
\end{figure}

\subsection{Phase diagram}

In this section we first review the ground-state properties of the model (\ref{H}) with disorder
by taking the limit of ${\mathbf A}=0$ in Eq.~(\ref{EilenFinGRn}) and considering a uniform system, $\bm{\nabla}\mathcal{G}^{(0)}_\omega=0$. We thus have
\begin{equation}\label{Eq4Eilen3}
\left[\omega_n\hat{\tau}_3\hat{\rho}_3\hat{\sigma}_0,\hat{\cal G}^{(0)}_\omega\right]+
i\left[(\hat{H}_{\textrm{mf}}+\hat{\Sigma}_\omega)\cdot\hat{\tau}_3\hat{\rho}_3\hat{\sigma}_0,\hat{\cal G}^{(0)}_\omega\right]=0.
\end{equation}
Without loss of generality we choose the SDW magnetization to be along the $z$-axis, $\bm{M}=M\bm{e}_z$. Then, the solution
of (\ref{Eq4Eilen3}) has the following form:
\begin{equation}\label{Ansatz4G}
\hat{\cal G}^{(0)}_\omega=g_{\omega}\hat{\tau}_3\hat{\rho}_3\hat{\sigma}_0-if_{\omega}\hat{\tau}_0\hat{\rho}_1\hat{\sigma}_2-is_{\omega}\hat{\tau}_2\hat{\rho}_0\hat{\sigma}_3,
\end{equation}
where the functions $g_{\omega}$, $f_{\omega}$, and $s_{\omega}$ are determined by the solution of the following system of algebraic equations:
\begin{equation}\label{FinEilen}
\begin{split}
i\Delta g_{\omega}&=f_{\omega}\left(\omega_n+2\Gamma_{\pi}g_{\omega}\right),\\
iMg_{\omega}&=s_{\omega}\left(\omega_n+2\Gamma_t g_{\omega}\right).
\end{split}
\end{equation}
Here we introduced the total scattering rate $\Gamma_t=\Gamma_0+\Gamma_\pi$. In addition, the functions  $g_{\omega}$, $f_{\omega}$, and $s_{\omega}$ satisfy the normalization condition $g_{\omega}^2-f_{\omega}^2-s_{\omega}^2=1$. Subsequently, superconducting and SDW order parameters can be found from
\begin{equation}\label{selfEilen}
\frac{iM}{g_m}=\pi T\sum\limits_{\omega_n>0}^\Lambda s_{\omega}, \quad
\frac{i\Delta}{g_{sc}}=\pi T\sum\limits_{\omega_n>0}^\Lambda f_{\omega},
\end{equation}
where $g_{sc}$ and $g_m$ are the coupling constants, and $\Lambda$ is an ultraviolet cutoff. In the clean system, there is a phase transition from the paramagnetic to the SDW state at critical temperature $T_{s0}=1.13\Lambda e^{-2/\nu g_m}$ provided that $g_{m}>g_{sc}$. If $g_{sc}>g_{m}$, the ground state is a superconductor with a critical temperature $T_{c0}=1.13\Lambda e^{-2/\nu g_{sc}}$. We consider $T_{s0}>T_{c0}$, so that without disorder, the SDW phase develops at a higher temperature.

\begin{figure}
\includegraphics[width=1\linewidth]{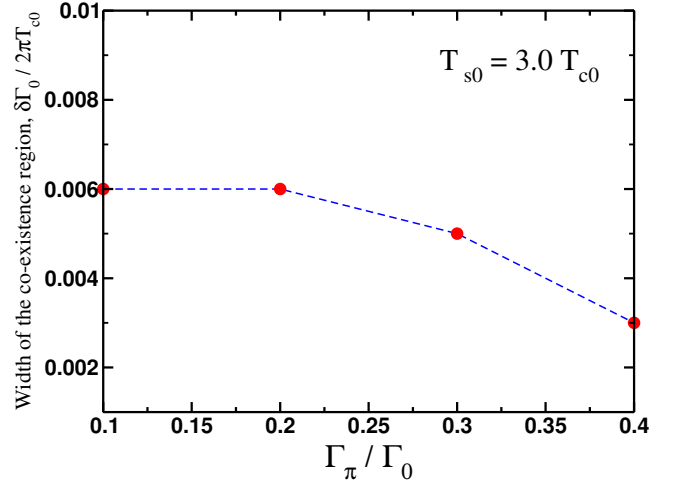}
\caption{\label{Fig2Width} (Color online) Width of the coexistence region $(\Gamma_{0}^{(\textrm{sdw})}-\Gamma_{0}^{(\mathrm{sc})})/2\pi T_{c0}$ is shown as function of $\Gamma_\pi/\Gamma_0$. The data points are found from the solution of Eqns.~(\ref{selfEilen}) and (\ref{FinEilen}) at zero temperature. The plot is obtained assuming $T_{s0}=3T_{c0}$.}
\end{figure}

\begin{figure}
\includegraphics[width=0.8\linewidth]{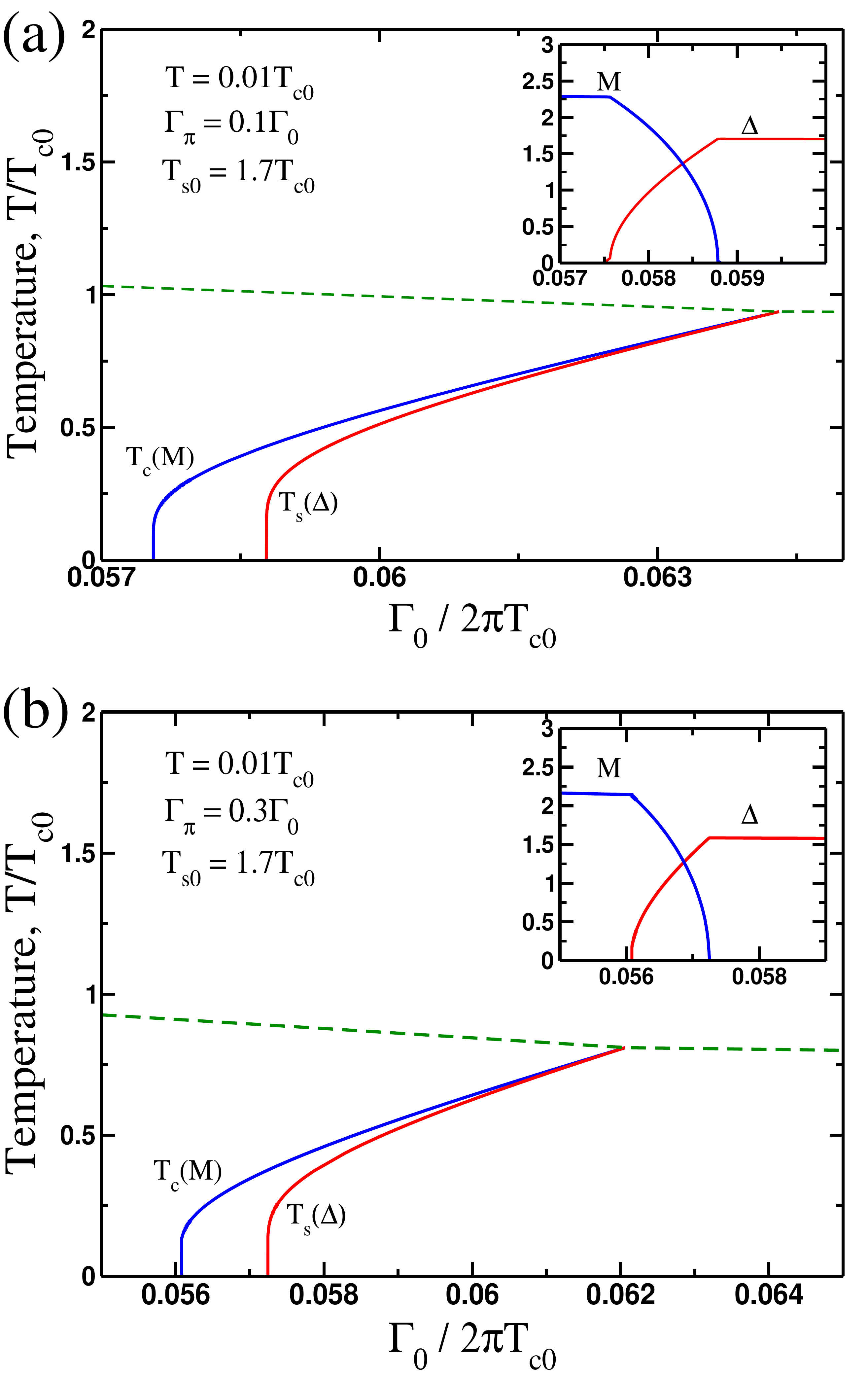}
\caption{\label{Fig3NewDia} (Color online) Phase diagrams in the $(\Gamma_0,T)$ plane for $T_{s0}=1.7T_{c0}$. In panel (a) we show the variation of the critical temperatures with $\Gamma_0$ for $\Gamma_\pi=0.1\Gamma_0$, while panel (b) shows critical temperatures for $\Gamma_\pi=0.3\Gamma_0$. The insets show the variation of the pairing amplitude and magnetization with disorder, evaluated at $T=0.01T_{c0}$. We use these data to evaluate the variation of the London penetration depth with disorder and temperature.}
\end{figure}

We solve Eqs.~(\ref{FinEilen}) and (\ref{selfEilen}) numerically, and show our results in Figs.~\ref{Fig1Dia}-\ref{Fig3NewDia}. In agreement with an earlier work,~\cite{VC-PRB11} we find that for a narrow region in $\Gamma_0$ values, there is a region in the phase diagram where SDW and superconductivity coexist. Specifically, superconductivity emerges when $\Gamma_0$ reaches some value denoted by $\Gamma_{0}^{(\mathrm{sc})}$. With further increase of intraband scattering, the SDW order is fully suppressed at some value $\Gamma_0=\Gamma_{0}^{(\textrm{sdw})}$. As the ratio $\Gamma_\pi/\Gamma_0$ increases, both $\Gamma^{(\mathrm{sc})}_{0}$ and $\Gamma_{0}^{(\textrm{sdw})}$ decrease. In Fig.~\ref{Fig2Width} we plot the width of the coexistence region $(\Gamma_{0}^{(\textrm{sdw})}-\Gamma_{0}^{(\mathrm{sc})})/2\pi T_{c0}$ at $T=0$. Thus, we conclude that the coexistence region remains quite robust with respect to the interband scattering, and it only vanishes when both scattering rates become comparable, $\Gamma_0\sim\Gamma_\pi$.

\section{London penetration depth}

In this section we solve the Eilenberger equation (\ref{EilenFinGRn}) and use the resulting correction to the Eilenberger function to compute the penetration depth as a function of disorder and temperature in the lowest order in $\mathbf{A}$.

\subsection{Solution of the Eilenberger equation in an external magnetic field}

The field-induced correction to the Eilenberger function (\ref{Gperturb}) is given by the solution of the following matrix equation
\begin{equation}\label{EilenFinGRn2}
\begin{split}
&\left[i\omega_n\hat{\tau}_3\hat{\rho}_3\hat{\sigma}_0,\hat{\cal G}_\omega^{(1)}({\mathbf R},{\mathbf n})\right]-\left[\hat{H}_{\mathrm{mf}}\hat{\tau}_3\hat{\rho}_3\hat{\sigma}_0,\hat{\cal G}_\omega^{(1)}({\mathbf R},{\mathbf n})\right]\\
&
-\left[\hat{\Sigma}_{\omega}\hat{\tau}_3\hat{\rho}_3\hat{\sigma}_0,\hat{\cal G}_\omega^{(1)}({\mathbf R},{\mathbf n})\right]=-\left[ev_F\bn\cdot{\mathbf A}\hat{\tau}_0\hat{\rho}_3\hat{\sigma}_0,\hat{\cal G}^{(0)}_\omega\right],
\end{split}
\end{equation}
which is found from (\ref{EilenFinGRn}) by keeping terms linear in the vector potential. The function $\hat{\cal G}_\omega^{(1)}({\mathbf R},{\mathbf n})$ must also satisfy the following condition, which results from the normalization of the full Eilenberger function (\ref{Gperturb}):
\begin{equation}\label{NewNorm}
\hat{\cal G}^{(0)}_\omega\cdot\hat{\cal G}_\omega^{(1)}({\mathbf R},{\mathbf n})+\hat{\cal G}_\omega^{(1)}({\mathbf R},{\mathbf n})\cdot\hat{\cal G}^{(0)}_\omega=0.
\end{equation}
We look for a solution of this equation in the following form
\begin{equation}\label{G1Anzats}
\begin{split}
\hat{\cal G}_\omega^{(1)}(\bR,\bn)&=g_{\omega}^{(1)}(\bR,\bn)\hat{\tau}_0\hat{\rho}_3\hat{\sigma}_0
-if_{\omega}^{(1)}(\bR,\bn)\hat{\tau}_3\hat{\rho}_1\hat{\sigma}_2\\&-\hat{s}_{\omega}^{(1)}(\bR,\bn).
\end{split}
\end{equation}
The matrix form for the first two terms follows from solving Eq.~(\ref{EilenFinGRn2}); first in the limit when $\Delta=M=0$, and then for $M=0$. In order to find the matrix structure of the third term we use condition (\ref{NewNorm}), which can only be fulfilled for
\begin{equation}\label{FindS1}
\hat{s}_{\omega}^{(1)}(\bR,\bn)=s_{\omega}^{(1)}\hat{\tau}_2\hat{\rho}_2\hat{\sigma}_1.
\end{equation}
Using condition (\ref{NewNorm}) we obtain for the functions $g_{\omega}^{(1)}$, $f_{\omega}^{(1)}$, and $s_{\omega}^{(1)}$ :
\begin{equation}\label{Quest4S1}
\begin{split}
&g_{\omega}^{(1)}(\bR,\bn)=-\frac{f_{\omega}^2}{z_\omega}{ev_F\mathbf n}\cdot{\mathbf A}(\bR), \\
&f_{\omega}^{(1)}(\bR,\bn)=-\frac{f_{\omega}g_{\omega}}{z_{\omega}}{ev_F\mathbf n}\cdot{\mathbf A}(\bR), \\
&s_{\omega}^{(1)}(\bR,\bn)=-\frac{s_{\omega}g_{\omega}}{z_\omega}{ev_F\mathbf n}\cdot{\mathbf A}(\bR),
\end{split}
\end{equation}
with
\begin{equation}\label{Zw}
\begin{split}
z_\omega&=(i\omega_n+i\Gamma_tg_{\omega})g_{\omega}+(\Delta-i\Gamma_sf_{\omega})f_{\omega}\\&+(M+i\Gamma_ts_{\omega})s_{\omega},
\end{split}
\end{equation}
where $\Gamma_s=\Gamma_0-\Gamma_\pi$. Equations (\ref{Quest4S1}) and (\ref{Zw}) constitute the perturbative solution of the Eilenberger equation (\ref{EilenFinGRn}).

\begin{figure}
\includegraphics[width=0.85\linewidth]{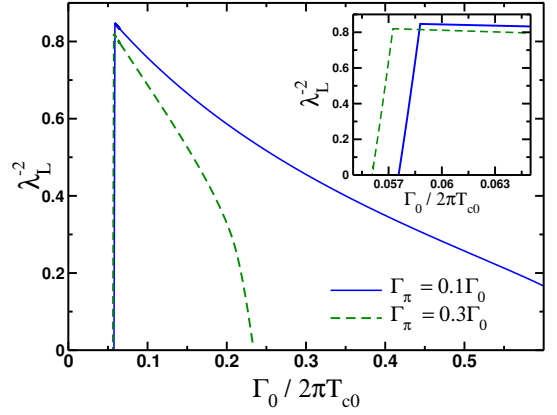}
\caption{\label{Fig4London} (Color online) Inverse square of the London penetration depth (dimensionless units) as a function of intraband scattering rate $\Gamma_0$ evaluated at $T=0.01T_{c0}$. In the co-existence region $\lambda_L^{-2}$ grows linearly with $\Delta$ in contrast with the clean case where $\lambda_L^{-2}\sim\Delta^2$, Eq.~(\ref{iLamL2Clean}).}
\end{figure}

\begin{figure}
\includegraphics[width=0.85\linewidth]{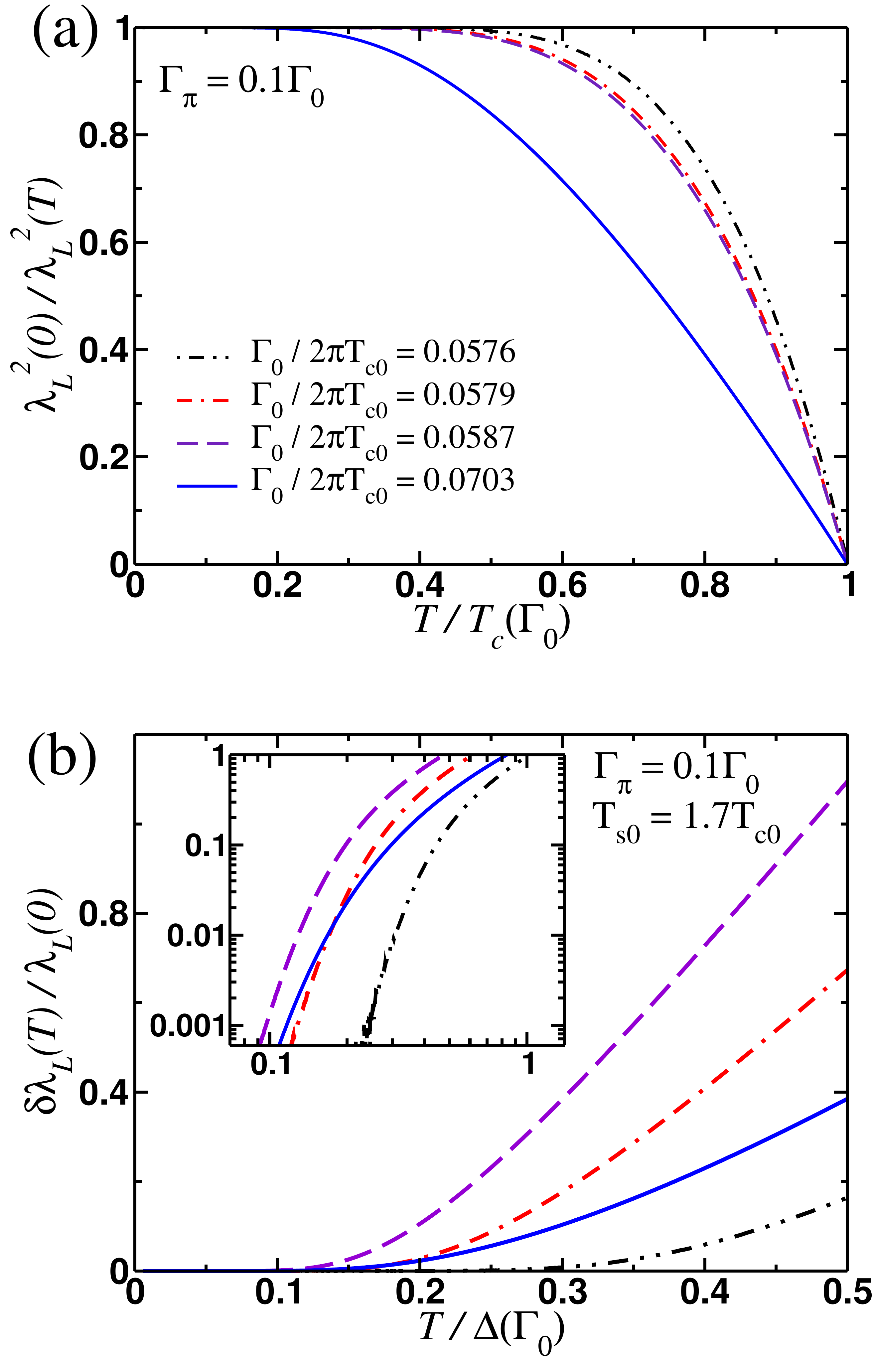}
\caption{\label{Fig5Ruslan} (Color online) a) Temperature dependence of $\lambda_L^{-2}$ for various values of $\Gamma_0$.
b) Temperature dependence for $\delta\lambda_L(T)=\lambda_L(T)-\lambda_L(0)$ at low temperatures. The inset shows the same data as in the main panel-(b) but plotted in log-log scale. 
The interband scattering rate is fixed at $\Gamma_\pi=0.1\Gamma_0$.}
\end{figure}

\subsection{Effect of disorder on the penetration depth}

The expression for the current density in terms of the original Green's function is given by\cite{AGD}
\begin{equation}\label{Jrfin2}
\begin{split}
\bm{j}=&\frac{eT}{2m}\sum\limits_{\omega_n}\int\limits\frac{d^2\bp}{(2\pi)^2}\bp\textrm{Tr}
\left[\hat{\tau}_3\hat{\rho}_0\hat{\sigma}_0\hat{G}_{\omega}^{(1)}(\bR,\bp)\right]\\&-\frac{Ne^2\mathbf{A}}{mc},
\end{split}
\end{equation}
where the last term guarantees the gauge invariance of the normal state. We can now use the same approximation that we have already employed in our derivation of the Eilenberger equation: since the main contribution to the current comes from a narrow energy region around the Fermi surface, in the integral above we approximate $\bp\approx p_F\bn$. It follows that
\begin{equation}\label{Jrfin4}
\begin{split}
\bm{j}=\frac{-ie\pi \nu p_FT}{8m}\sum\limits_{\omega_n}\left\langle\bn\textrm{Tr}
\left[\hat{\tau}_0\hat{\rho}_3\hat{\sigma}_0\hat{\cal G}_{\omega}^{(1)}(\bR,\bn)\right]\right\rangle,
\end{split}
\end{equation}
where we used the definition (\ref{EilenG}); angular brackets denote averaging over directions of unit vector ${\mathbf n}$. Using our solution of the Eilenberger equation together with the equations (\ref{FinEilen}) for the current we obtain at an intermediate step
\begin{equation}
\bm{j}=-Q\mathbf{A},\qquad Q=-\frac{\nu e^2v^2_F}{2c}T\sum_{\omega_n}\frac{if^2_\omega}{z_\omega},
\end{equation}
where the auxiliary function $z_\omega$ is defined by Eq.~(\ref{Zw}).
This expression can be significantly simplified by using the mean-field equations (\ref{FinEilen}) together with the normalization condition. Indeed, one observes that first and third terms in $z_\omega$ can be combined as follows
\begin{equation}
\begin{split}
&(\omega_n+\Gamma_tg_\omega)g_\omega-i(M+i\Gamma_ts_\omega)s_\omega=\\
&(\omega_n+\Gamma_tg_\omega)g_\omega-(iMg_\omega-\Gamma_ts_\omega g_\omega)\frac{s_\omega}{g_\omega}=\\
&(\omega_n+\Gamma_tg_\omega)g_\omega-(\omega_n+\Gamma_tg_\omega)\frac{s^2_\omega}{g_\omega}=(\omega_n+\Gamma_tg_\omega)\frac{1+f^2_\omega}{g_\omega},
\end{split}
\end{equation}
and then $(\omega_n+\Gamma_tg_\omega)(1+f^2_\omega)-i(\Delta-i\Gamma_sf_\omega)f_\omega g_\omega=(\omega_n+\Gamma_tg_\omega)$. Consequently $Q$ can be brought to the form
\begin{equation}\label{jRfin}
Q=\frac{\nu e^2v_F^2}{2c}T\sum\limits_{\omega_n}\frac{f_{\omega}^2g_{\omega}}{\omega_n+\Gamma_t g_{\omega}}.
\end{equation}
Curiously, the function $s_{\omega}$ does not enter explicitly into the final expression for the current. Thus, for the London penetration depth we have
\begin{equation}\label{iLamL2}
\lambda_L^{-2}(T)=\frac{\nu e^2v_F^2}{c^2}2\pi T\sum\limits_{\omega_n}\frac{f_{\omega}^2g_{\omega}}{\omega_n+\Gamma_t g_{\omega}},
\end{equation}
which is the main result of this paper. Next we analyze various limiting cases.

In the clean limit, $\Gamma_t=0$, it is easy to show that
\begin{equation}\label{iLamL2Clean}
\lambda^{-2}_L=\lambda^{-2}_L(0)\frac{\Delta^2}{M^2+\Delta^2},
\end{equation}
in agreement with earlier studies.~\cite{Fernandes,Kuzmanovski} Let us now analyze the Matsubara sum in (\ref{iLamL2}) in the limit of low temperatures and for weak interband disorder, $\Gamma_\pi\ll\Gamma_0$. In that limit, using Eqns.~(\ref{FinEilen}) for the function $f_\omega$, we find $f_\omega\approx{i\Delta g_\omega}/{\omega}$.
Next, we set $T\to 0$, and convert the frequency summation into an integral over the variable $x=\omega/\Delta$. The resulting expression for the Matsubara sum in (\ref{iLamL2}) has the following form
\begin{equation}\label{Int3}
\begin{split}
2\pi T\sum_\omega\to\int_{0}^{\infty}\frac{x[x+\gamma g(x)]^{-1}dx}{\left(x^2+1+\frac{M^2x^2}{\Delta^2[x+2\gamma g(x)]^2}\right)^{3/2}},
\end{split}
\end{equation}
where we have introduced the parameter $\gamma={\Gamma_t}/{\Delta}$. For small enough values of $\Delta$, such that $\gamma\gg 1$, it follows that for the moderate range of $x\sim O(1)$ we can simplify $x+2\gamma g(x)\approx2\gamma g(x)$.
Furthermore, the dominant contribution to the integral (\ref{Int3}) comes from the region of $x$ where $g(x)\sim x$.
Thus, for sufficiently small $\Delta$, we approximately obtain for the integral
\begin{equation}\label{approxLam}
\lambda^{-2}_L\simeq\lambda^{-2}_L(0)\frac{\Delta}{\sqrt{M^2+4\Gamma_0^2}}.
\end{equation}
In the opposite limit, and still at zero temperature, another analytical result for $\lambda_L$ as a function of the scattering rates and $\Delta$ can be derived (see Appendix B for details)
\begin{equation}\label{IM0res}
\begin{split}
&\lambda^{-2}_L=\lambda^{-2}(0)\left[\frac{\pi(\gamma_\pi+2\gamma_\pi\gamma_s^2+4\gamma_s^3)}{12\gamma_s^4}\right. \\
&\left.-\frac{\gamma_\pi(3+8\gamma_s^2)}{12\gamma_s^3}-\frac{(\gamma_\pi+4\gamma_s^3)\arccos(2\gamma_s)}{8\gamma_s^4\sqrt{1-4\gamma_s^2}}\right],
\end{split}
\end{equation}
where we introduced the following parameters for brevity $\gamma_s=2\Gamma_s/\Delta$, and $\gamma_\pi=2\Gamma_\pi/\Delta$. Note that when interband scattering becomes negligibly small, then the term of Eq.~(\ref{IM0res}) in square brackets is proportional to $\Delta/\Gamma_0$, in agreement with our estimate, Eq.~(\ref{approxLam}), taken in the same limit.

Our complete numerical analysis of equation (\ref{iLamL2}) confirms our asymptotic analytical expressions, and in particular the estimate (\ref{approxLam}). In fact, we find that this behavior persists for much larger values of $\Delta\sim \Gamma_t$. Lastly, comparing this result with the corresponding expression for the London penetration depth in the clean limit (\ref{iLamL2Clean}), we conclude that disorder has a crucial effect on the dependence of $\lambda_L^{-2}$ on both $\Delta$ and $M$.

\begin{figure}[t!]
\includegraphics[width=0.85\linewidth]{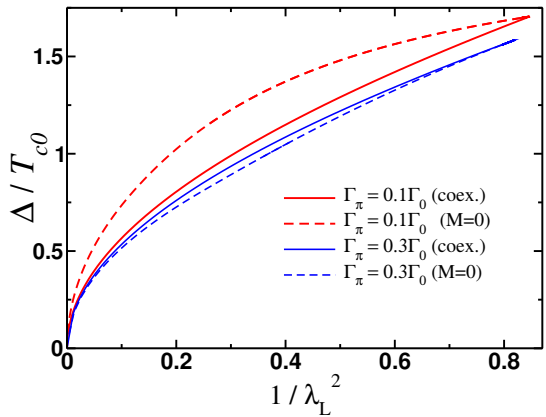}
\caption{\label{Fig6Ophir} (Color online) Plot of $\lambda_L^{-2}$ on $\Delta$ across the phase diagram at $T=0.01T_{c0}$. Remarkably, the position of the line for the purely superconducting state depends on the strength of the interband scattering rate. This is expected as the superconducting order parameter is strongly suppressed for higher values of $\Gamma_\pi/\Gamma_0$. }
\end{figure}

In Fig.~\ref{Fig4London} we show the variation of $\lambda_L^{-2}$ with disorder, computed using Eq.~(\ref{iLamL2}) together with $\Delta(\Gamma_0)$ and $M(\Gamma_0)$, which in turn have been computed self-consistently and are shown in Fig.~\ref{Fig3NewDia}. Furthermore, in Fig.~\ref{Fig5Ruslan} we show the temperature dependence of $\lambda_L^{-2}(T)$ for various values of $\Gamma_0$ across the phase diagram [see inset (a) in Fig.~\ref{Fig3NewDia}]. Finally, in Fig.~\ref{Fig6Ophir} we show the dependence of $\Delta$ on $\lambda_L^{-2}$ evaluated at $T=0.01T_{c0}$. One immediately observes that combining measurements of the London penetration depth in the coexistence region with those made in the superconducting state should, in principle, allow one to obtain an estimate for the ratio between the interband and intraband scattering rates. Indeed, for moderate values of $\Gamma_\pi/\Gamma_0$, superconductivity is strongly suppressed, leading to lower values of $\Delta$ in the superconducting state compared to those in the coexistence state.

\section{Discussions and Perspectives}

In this paper, we obtained the phase diagram of doped iron-pnictide superconductors and calculated the magnetic penetration depth at different temperatures under the assumption that doping introduces disorder but does not affect the band structure. In several limiting cases we have been able to reproduce previously known results. 

Our main finding concerns the behavior of $\lambda_L$ in the coexistence phase that has not been systematically analyzed before in the presence of disorder scattering. Our modeling shows that starting from the overdoped side, $\lambda^{-2}_L$ grows with the reduction of the scattering rate induced by doping up to an optimal doping where superconducting order is maximal. Once the system enters the coexistence phase, the inverse square of the penetration depth, which is proportional to the superfluid density, exhibits a kink followed by a sharp falloff. In that region, disorder primarily affects magnetic order rather than superconductivity. In sharp contrast with the clean case, where $\lambda^{-2}_L\propto\Delta^2$ near the end-point of the superconducting dome from the side of the pre-existing SDW, within the disorder model we find a completely different scaling law $\lambda^{-2}_L\propto\Delta$. Another important observation concerns the low temperature dependence of $\delta\lambda_L(T)$. The log-log plot presented in Fig.~(\ref{Fig5Ruslan}b) suggests either a power law behavior of $\delta\lambda_L$ on temperature, $\delta\lambda_L\propto T^a$, with rather high power exponent $a\gtrsim 5$, or exponential dependence $\delta\lambda_L\propto e^{-W/T}$. The exponential dependence indicates the presence of the gap $W$ in the spectrum of electron states near the Fermi surface for parameters of curves presented in Fig.~(\ref{Fig5Ruslan}).  It is plausible, however, that for stronger $\Gamma_\pi$ electron spectrum becomes gapless at low temperature and $\delta \lambda_L(T)$ exhibits a power law with $a\simeq 2$.~\cite{VVC-PRB09}

Several extensions of the presented model are in order to improve the comparison with experiments. First, one could treat the band~\cite{VVC-PRB10,FS-PRB10} and disorder~\cite{VC-PRB11,FVC} models on an equal footing in order to study the observed anisotropy in $\lambda_L$. Second, one could consider the extension of the presented formalism beyond the Born approximation, which might be necessary for an accurate interpretation of the low-temperature data. Third, one could account for diffusive scattering from the surface of the superconductor. That would complicate the calculation of the penetration length considerably, since one would have to work with an integral Milne equation, instead of a differential London equation, which governs the distribution of the magnetic field in superconductors and consequently determines the precise value of $\lambda_L$ in the nonlocal limit.~\cite{Miller} Fourth, our analysis, so far, has been restricted to the mean-field level. A most intriguing recent experimental observation~\cite{Matsuda,Auslaender} is an apparent sharp peak in $\lambda_L$ observed in isovalently P-doped BaFe$_2$As$_2$ at nearly zero temperature around the optimal doping. This effect was attributed to quantum critical fluctuations of the SDW order at the onset of the transition into the coexistence phase.~\cite{AL,Chowdhury,Nomoto-Ikeda} It is of apparent theoretical and experimental necessity to investigate to what degree such quantum effects are robust against disorder scattering. On a technical level, this would require the inclusion of magnetization fluctuations into the existing formalism. We note that such a generalization has already been carried out in the context of thermal magnetic fluctuations, which are relevant for the interpretation of specific heat data.~\cite{Dushko} Finally, when analyzing the quantum critical behavior of the superfluid density in FeSC compounds, it might be useful to use the results obtained in the context of cuprate superconductors.~\cite{Randeria} In the vicinity of the quantum critical point (QCP), generic scaling analysis indicates that the superconducting critical temperature should vanish as $T_c\propto \delta^{z\nu}$, where $\delta=|x-x_c|$ measures the deviation in doping from the QCP, while $z$ and $\nu$ are the quantum dynamical and correlation length exponents. At the same time, the superfluid density should scale as $n_s\propto\lambda^{-2}_L\propto \delta^{(z+d-2)\nu}$, where $d$ is the dimensionality of the system. When combined together, these two scaling laws predict that there should exist a precise relation $d\ln T_c/d\ln n_s=z/(z+d-2)$. In the two-dimensional case there should exist a linear relation between $T_c$ and $n_s$. For FeSCs, each of the end-points of the coexistence phase represent a QCP and, consequently, establishing the relation between $T_c$ and $n_s$ will provide new information about superconductivity in these complex materials.

\section*{Acknowledgments}

We would like to thank O. Auslaender, R. Fernandes, D. Kuzmanovski, Y. Matsuda, R. Prozorov, J. Schmalian, and T. Shibauchi for fruitful discussions. This work was supported by NSF Grant DMR-1401908 and in part by DAAD grant from German Academic Exchange Services (A.L.), by KSU, MPI-PKS and NSF Grant No. DMR-1506547 (M.D.), and NSF Grant No. DMR-0955500 (M.V.). M.K and A.L. acknowledge the support from the Binational Science Foundation Grant No. 2014107.

\begin{appendix}

\section{Derivation of the Eilenberger equation}
We start by writing down the equations of motion for the matrix Green's function using an imaginary time representation, while ignoring the disorder potential for the time being:
\begin{eqnarray}\label{EQ4Gr1r2}
-\frac{\partial}{\partial\tau_1}\hat{G}(\br_1\tau_1;\br_2\tau_2)-\hat{H}(\br_1)\cdot\hat{G}(\br_1\tau;\br_2\tau_2)\nonumber\\ =\hat{I}\delta(\tau_1-\tau_2)\delta(\br_1-\br_2),\\
\frac{\partial}{\partial{\tau_2}}\hat{G}(\br_1\tau_1;\br_2\tau_2)-\hat{G}(\br_1\tau_1;\br_2\tau_2)\cdot\hat{H}(\br_2)\nonumber\\=\hat{I}\delta(\tau_1-\tau_2)\delta(\br_1-\br_2).
\end{eqnarray}
Next, we use mixed space-momentum, keeping the center-of-mass coordinate $\bR=(\br_1+\br_2)/2$, but making the Fourier transformation with respect to the relative coordinate $\br=\br_1-\br_2$. In the Matsubara frequency representation, we find:
\begin{eqnarray}\label{EilenSystemMain-1}
&&\hskip-.85cm
\left[i\omega_n\hat{\tau}_3\hat{\rho}_3\hat{\sigma}_0+\xi_\bp
-\frac{i\bp\cdot{\mbox{\boldmath $\nabla$}_{\mathbf R}}}{2m}\right]\hat{\tau}_3\hat{\rho}_3\hat{\sigma}_0\hat{G}_\omega({\mathbf R},\bp)\nonumber\\
&&\hskip-.85cm
+{\mathbf v}\cdot{\mathbf A}({\mathbf R})\hat{\tau}_3\hat{\rho}_0\hat{\sigma}_0\hat{G}_\omega({\mathbf R},\bp)-\hat{H}_{\mathrm{mf}}({\mathbf R})\hat{G}_\omega({\mathbf R},\bp)=\hat{I},\\
&&\hskip-.85cm
i\omega_n\hat{G}_\omega({\mathbf R},\bp)+\left(\xi_\bp+\frac{i\bp\cdot{\mbox{\boldmath $\nabla$}_{\mathbf R}}}{2m}\right)\hat{G}_\omega({\mathbf R},\bp)\hat{\tau}_3\hat{\rho}_3\hat{\sigma}_0+\nonumber\\
&&\hskip-.85cm
\hat{G}_\omega({\mathbf R},\bp)\hat{\tau}_3\hat{\rho}_0\hat{\sigma}_0
{\mathbf v}\cdot{\mathbf A}({\mathbf R})-\hat{G}_\omega({\mathbf R},\bp)\hat{H}_{\mathrm{mf}}({\mathbf R})=\hat{I},\label{EilenSystemMain-2}
\end{eqnarray}
where we redefined the vector potential $\frac{e}{c}{\mathbf A}\to{\mathbf A}$ for brevity, and we used Eqns.~(\ref{H0ab}) and (\ref{Hmf}). We write $\frac{\bp}{m}\approx v_F{\mathbf n}, \quad {\mathbf v}\approx v_F\bn, \quad {\mathbf n}=\frac{\bp}{p}$, and using Eq.~(\ref{EilenG}) we multiply Eq.~(\ref{EilenSystemMain-1}) by $\hat{\tau}_3\hat{\rho}_3\hat{\sigma}_0$ from the left, and multiply Eq.~(\ref{EilenSystemMain-2}) by the same matrix from the right. Then, we subtract the second equation from the first. Lastly, we use $\hat{\tau}_3\hat{\rho}_3\hat{\sigma}_0\cdot\hat{\tau}_3\hat{\rho}_3\hat{\sigma}_0=\hat{I}$ where necessary, and integrate both parts over the absolute value of the momentum, which allows us to use the Eilenberger Green's function (\ref{EilenG}). We thus find
\begin{eqnarray}
&&\hskip-.45cm
i\omega_n\hat{\cal G}_\omega({\mathbf R},{\mathbf n})-i\omega_n\hat{\tau}_3\hat{\rho}_3\hat{\sigma}_0\hat{\cal G}_\omega({\mathbf R},{\mathbf n})\hat{\tau}_3\hat{\rho}_3\hat{\sigma}_0\nonumber\\
&&\hskip-.45cm
+v_F\bn\cdot(-i{\mbox{\boldmath $\nabla$}_\bR})
\hat{\tau}_3\hat{\rho}_3\hat{\sigma}_0\hat{\cal G}_\omega({\mathbf R},{\mathbf n})\nonumber\\
&&\hskip-.45cm
+v_F\bn\cdot{\mathbf A}(\bR)
\left[\hat{\tau}_3\hat{\rho}_0\hat{\sigma}_0\hat{\cal G}_\omega({\mathbf R},{\mathbf n})-\hat{\tau}_3\hat{\rho}_3\hat{\sigma}_0\hat{\cal G}_\omega({\mathbf R},{\mathbf n})\hat{\tau}_0\hat{\rho}_3\hat{\sigma}_0\right]\nonumber\\
&&\hskip-.45cm
+\hat{\tau}_3\hat{\rho}_3\hat{\sigma}_0\hat{\cal G}_\omega({\mathbf R},{\mathbf n})\hat{H}_{\mathrm{mf}}({\mathbf R})\hat{\tau}_3\hat{\rho}_3\hat{\sigma}_0-\nonumber\\
&&\hskip-.45cm
-\hat{\tau}_3\hat{\rho}_3\hat{\sigma}_0
\hat{H}_{\mathrm{mf}}({\mathbf R})\hat{\tau}_3\hat{\rho}_3\hat{\sigma}_0\hat{\cal G}_\omega({\mathbf R},{\mathbf n})=0.
\end{eqnarray}
This equation can be written in a compact form if we multiply it from the left by $\hat{\tau}_3\hat{\rho}_3\hat{\sigma}_0$. Then, Eq.~(\ref{EilenFinGRn}) from the main text follows,
where we have included the effects of disorder by trivially writing the self-energy correction to the mean-field Hamiltonian. In terms of the Eilenberger Green's function it is given by
\begin{eqnarray}\label{SigmaR}
&&\hat{\Sigma}_{\omega}(\bR)=-i\Gamma_0\hat{\tau}_0\hat{\rho}_3\hat{\sigma}_0\hat{\tau}_3\hat{\rho}_3\hat{\sigma}_0
\int\frac{d\phi_\bn}{2\pi}\hat{\cal G}_{\omega}(\bR,\bn)\hat{\tau}_0\hat{\rho}_3\hat{\sigma}_0\nonumber\\
&&-i\Gamma_\pi\hat{\tau}_1\hat{\rho}_3\hat{\sigma}_0
\hat{\tau}_3\hat{\rho}_3\hat{\sigma}_0\int\frac{d\phi_\bn}{2\pi}\hat{\cal G}_{\omega}(\bR,\bn)\hat{\tau}_1\hat{\rho}_3\hat{\sigma}_0.
\end{eqnarray}

\section{London penetration depth for $M\ll \Delta$}
In this section we will derive an expression for the London penetration depth at low temperatures, assuming that the SDW order parameter is much smaller than the superconducting order parameter, $M\ll \Delta$. We start by writing Eqns.~(\ref{FinEilen})
where we replace $f\to if$ and $s\to is$:
\begin{equation}\label{fifSiS}
\Delta g_\omega=(\omega_n+2\Gamma_\pi)f_\omega, \quad M g_\omega=(\omega_n+2\Gamma_t)s_\omega.
\end{equation}
The functions in (\ref{fifSiS}) satisfy the normalization condition $g_\omega^2+f_\omega^2+s_\omega^2=1$. From Eqs.~(\ref{fifSiS}) it follows that
\begin{equation}\label{SF}
\Delta s_\omega-Mf_\omega=-2\Gamma_0f_\omega s_\omega.
\end{equation}
We can now eliminate $s_\omega$ from this equation by using the normalization condition, which yields the following equation for $g_\omega$:
\begin{equation}
\frac{1}{1-g_\omega^2-f_\omega^2}\left[1-\frac{\Delta}{Mf_\omega}\sqrt{1-g_\omega^2-f_\omega^2}\right]^2=\frac{4\Gamma_0^2}{M^2}.
\end{equation}
Solving this equation for $g_\omega^2$ one obtains
\begin{equation}\label{gomega}
g_\omega^2=1-f_\omega^2-\frac{M^2f_\omega^2}{(2\Gamma_0f_\omega+\Delta)^2}.
\end{equation}
Next, we consider the following integral
\begin{equation}\label{IDM}
Q(\Delta,M)=-\int_{0}^{\infty}\frac{f_\omega^2g_\omega d\omega}{\omega+\Gamma_tg_\omega},
\end{equation}
which in a way determines the penetration depth. The idea is to replace the integration over $\omega$ with an integral over $f_\omega$. To do that, we employ Eqs.~(\ref{fifSiS}), (\ref{SF}), and (\ref{gomega}). It follows then that
\begin{equation}\label{IDM2}
\begin{split}
&Q(\Delta,M)=\int_{0}^{1}\frac{f_\omega df_\omega}{g_\omega(\Delta+\Gamma_sf_\omega)}\left[\Delta g_\omega^2\right.\\&
\left.+(\Delta-2\Gamma_\pi f_\omega)f_\omega^2\left(1+\frac{\Delta M^2}{(\Delta+2\Gamma_0 f_\omega)^3}\right)\right],
\end{split}
\end{equation}
and $g_\omega$ is a functional of $f_\omega$, Eq.~(\ref{gomega}). Clearly, for $\Gamma_t=0$ we find $Q(\Delta,M)=\Delta^2/(M^2+\Delta^2)$. When $M=0$, the expression for $Q(\Delta,M=0)$ simplifies to:
\begin{equation}\label{IM0}
Q(\Delta,0)=\int_{0}^{1}\frac{(\Delta-2\Gamma_\pi f^3)fdf}{\sqrt{1-f^2}(\Delta+2\Gamma_sf)}.
\end{equation}
This integral can be evaluated exactly, and it gives Eq.~(\ref{IM0res}) from the main text. Lastly, one can also expand $Q(\Delta,M)$ in powers of $M/\Delta$ to derive the correction to the penetration depth due to the development of the SDW order in the superconducting state. The resulting expression, however, is too cumbersome to show here.

\end{appendix}

\end{document}